\documentclass[12pt,preprint]{aastex}
\renewcommand {\deg}  {\mbox{$^\circ$}}

\newcommand  {\kms}  {\mbox{km\,s$^{-1}$}}
\begin{document}

%\title{ }
%\author{ }
%\email{ }
%\altaffiltext{1}{ }

%\begin{abstract}
%\vspace {5pt}
%\end{abstract}

%\keywords{ }
%\documentclass[12pt,preprint]{aastex}
%% manuscript produces a one-column, double-spaced document:
%% \documentclass[manuscript]{aastex}
%% preprint2 produces a double-column, single-spaced document:
%\documentclass[preprint2]{aastex}
%% Sometimes a paper's abstract is too long to fit on the
%% title page in preprint2 mode. When that is the case,
%% use the longabstract style option.
\def\kms {\hbox{km{\hskip0.1em}s$^{-1}$}} % km/s
\def\gos #1 {\left({G\over 10^3 G_0 }\right)^{#1}}
\def\ee #1 {\times 10^{#1}}          % \ee p      10^p
\def\msol{\hbox{$\hbox{M}_\odot$}}
\def\lsol{\hbox{$\hbox{L}_\odot$}}
\def\kms{km s$^{-1}$}
\def\Blos{B$_{\rm los}$}
\def\etal  {{\it et al. }}                    % et al
\def\psec          {$.\negthinspace^{s}$}
\def\pasec          {$.\negthinspace^{\prime\prime}$}
\def\pdeg          {$.\kern-.25em ^{^\circ}$}
\def\degree{\ifmmode{^\circ} \else{$^\circ$}\fi}
\def\ee #1 {\times 10^{#1}}          % \ee p      10^p
\def\ut #1 #2 { \, \textrm{#1}^{#2}} % \ut unit p  unit^p  
\def\u #1 { \, \textrm{#1}}          % \u unit    unit
\def\nH {n_\mathrm{H}}
\def\ddeg  {\hbox{$.\!\!^\circ$}}              % Degrees over dot
\def\deg    {$^{\circ}$}                        % Degrees symbol
\def\le    {$\leq$}                            % <=
\def\sec    {$^{\rm s}$}                        % Second of time
\def\msol  {\hbox{$M_\odot$}}                  % Solar mass
\def\i      {\hbox{\it I}}                      % italic I
\def\v      {\hbox{\it V}}                      % italic V 
\def\dasec  {\hbox{$.\!\!^{\prime\prime}$}}    % Arcseconds over dot
\def\asec  {$^{\prime\prime}$}                % Arcseconds symbol
\def\dasec  {\hbox{$.\!\!^{\prime\prime}$}}    % Arcseconds over dot
\def\dsec  {\hbox{$.\!\!^{\rm s}$}}            % Second over dot
\def\min    {$^{\rm m}$}                        % Minutes of time
\def\hour  {$^{\rm h}$}                        % Hours of time
\def\amin  {$^{\prime}$}                      % Arcminutes symbol
\def\lsol{\, \hbox{$\hbox{L}_\odot$}}
\def\sec    {$^{\rm s}$}                        % Second of time
\def\etal  {{\it et al. }}                    % et al.
\def\xbar  {\hbox{$\overline{\rm x}$}}        % bar over x

%\slugcomment{Submitted to ApJL}
%\shorttitle{structure function}
%\shortauthors{zadeh}
%\input psfig.sty

\title{Tidal Distortion  of the Envelope of an\\
AGB Star IRS 3 near Sgr A*}
\author{F. Yusef-Zadeh$^1$,  M. Wardle$^2$,  
W. Cotton$^3$, R. Sch\"odel$^4$, M. J. Royster$^1$, D. A. Roberts$^1$ \& D. Kunneriath$^3$ 
\\ 
}
\affil{$^1$Department of Physics \& Astronomy, Northwestern University, Evanston, IL 60208}
\affil{$^2$Department of Physics and Astronomy and
Research Center for Astronomy, Astrophysics\\
\& Astrophotonics, Macquarie University, Sydney NSW 2109, Australia}
\affil{$^3$National Radio Astronomy Observatory,  Charlottesville, VA 22903}
\affil{$^4$Instituto de Astfisica de Andalucia (CSIC),
Glorieta de la Astronomia S/N, 18008 Granada, Spain}

%and Center for Interdisciplinary Research in Astronomy,
%and Research Centre for Astronomy,
%Astrophysics \& Astrophotonics, 

\begin{abstract} 
We present radio and mm  continuum  observations  of the Galactic center taken with the VLA 
and ALMA at 44 and 226 GHz, respectively. We detect radio and mm emission from 
IRS 3, lying $\sim4.5''$ NW of Sgr A*, 
with  a spectrum that is consistent with the photospheric emission from an AGB star at the 
Galactic center. Millimeter images  reveal  that the envelope of 
IRS 3, the brightest and most extended 3.8$\mu$m Galactic center stellar source,  
consists of two  semi-circular  dust shells facing the direction of  Sgr A*. 
The outer circumstellar shell at the distance of  1.6$\times10^4$ AU, 
appears to break up into  ``fingers" of dust 
directed toward Sgr A*. These  features 
coincide with molecular CS (5-4) emission and a near-IR extinction cloud 
distributed between IRS 3 and Sgr A*. 
The NE-SW asymmetric shape  of the IRS  3 shells seen at  
 3.8$\mu$m  and radio
are   interpreted as structures that are 
tidally  distorted by Sgr A*. Using the kinematics of CS emission and the proper motion of IRS 3, 
the tidally distorted outflowing material from the envelope  after 5000 years
constrains  the distance  of IRS 3 to $\sim$0.7  pc in front of or $\sim$0.5 pc behind Sgr A*. 
This suggests that the mass loss  by stars near Sgr A* can supply 
a reservoir of molecular material near Sgr A*. 
We also present dark features in radio continuum images coincident with the 
envelope of IRS 3. 
These dusty stars  provide examples in which high resolution radio continuum images  
can  identify dust enshrouded stellar sources  embedded 
an ionized medium. 
\end{abstract} 

% and  the  M2I red supergiant star IRS 7.

%These dark features  coincide with cool dusty circumstellar shells 
%embedded in a hot ionized medium, tracing volumes 
%where there is a deficiency of free-free radio continuum emission. 

%We interpret these dark features as imprints of molecular gas and dust in the atmosphere of cool stars 


%\vfill\eject
%Apart from the G2 cloud, this is the first evidence of spatially  resolved 
%tidal distortion of the envelope of a star by Sgr A*.  

\section{Introduction}

The nucleus of our Galaxy hosts a supermassive black hole, Sgr~A*, at the dynamical center of the Galaxy. The stellar 
nuclear 
cluster surrounding Sgr~A* consists of a mixture of an evolved stellar population and a young population of stars at 
smaller radii. Sgr A* and its neighborhood are  subject to intense scrutiny, with potential for  
long-lasting impact on our understanding of massive black holes in the nuclei of normal galaxies. Understanding the 
processes occurring in the immediate environment of the massive black hole provides  insight into 
our own Milky Way's most massive black hole,
and presents an 
unparalleled opportunity to closely study the process by which gas is captured and radiated by supermassive black 
holes (SMBHs).

Recent OH, CN, CS, HCN and SiO observations  suggest that  molecular 
gas is able to survive  within 5$''$ (0.2 pc) 
of Sgr A* (Montero-Castano et 
al. 2009; Martin et al. 2009; Yusef-Zadeh et al. 2013; 
Moser et al. 2016; Karlsson et al. 2015). 
One key question is how this gas finds its way  so close to Sgr~A*. One 
possibility is the radial infall of giant molecular clouds 
toward the Galactic center.  
Another possibility is 
the highly eccentric orbit of compact clouds,  
such as  G2,  bring neutral  material close to  Sgr A*
(e.g., Gillessen et al. 2012, 2013). 
These clouds 
experience  tidal stripping and provide  the supply of accreting material  onto Sgr~A*. 

%which could have an embedded young stellar object (YSO) as its source,  that pass very close to Sgr~A* 

%There is no clear evidence for increased 
%activity of Sgr~A* at  pre- or  post-pericenter passage of the G2 cloud. 

IRS 3 is  the brightest and most extended 3.8$\mu$m star in the 
central pc of the Galaxy resembling  either a young massive star surrounded by dust 
(Krabbe et al. 2005; Tanner et al. 2005; Viehmann et al. 2005) or a cool dusty star (e.g., Roche and Aitken 1985; 
Pott et al. 2008). Recent high resolution 3.8$\mu$m  observations identified IRS 3 as a cool AGB star without any 
associated OH masers. 
The 3.8$\mu$m  emission has  two components, a compact and 
bright source, coincident with the central star with an effective stellar temperature of 3$\times10^3$K and 
 a dusty shell  with a radius of 1$''$ or (8000 AU)
surrounding the central star  (Pott et al. 2008).

%The dust shell of  this carbon rich star 
%is estimated to have  a dust temperature 1200K (Pott et al. 2008). 

%Near-IR observations have also detected a WR star IRS 3E, 0.12$''$ to the east of IRS 3 
%(Paumard et al. 2006). 

Here, we present ALMA and VLA observations of the Galactic 
center and show that the dusty outer shell of 
IRS 3 located $4.5''$ NW of Sgr~A* 
(projected distance
$\sim$0.18  pc at the Galactic center distance  8 kpc) 
is being tidally stretched by  Sgr~A*.  
The evidence for  tidal distortion of the envelope of IRS 3
implies  that  
dusty, evolved  stars with massive envelopes  approaching  the dynamical center of the Galaxy 
may   supply  the fuel for accretion  onto Sgr~A*.  

%This tidal stretching is likely responsible for 
%suppressing any maser emission.   

%If the increased mass loss of evolved stars due to tidal distortion 
%is significant, 
%this population of AGB stars may lose angular momentum and  spiral in toward the Galactic center.  

%IRS 3 is a  member of the population of evolved stars 
%with  a power law distribution away from Sgr~A* (e.g., Sch\"odel et al. 2007).

We also present new observations indicating multiple shells of dust  emission at mm 
extending up to $\sim$0.1 pc from the central star. We detect diffuse continuum radio 
emission  from the outermost shell of IRS 3 suggesting external photoionization 
by  the central young star cluster similar to IRS 7. 
In addition, we show several fingers of dust emission stretched from the outer shell of 
IRS 3  towards  Sgr~A* is located. Finally, 
dark features coincident with the dusty envelopes of IRS 3  
in radio continuum images are noted. 

%inspiraling of 
%and the loss of the angular momentum of 
%orbiting stars. 

%Alternatively, G2 could be a photo-evaporative wind from a proto-planetary disk
%surrounding a low-mass star (e.g., Murray-Clay \& Loeb 2012). 
%, thus suggesting 
%a  star may be embedded  in G2.

%In spite of its large extent at mid-IR,  there has not been 
%any detection of radio  continuum emission from the stellar envelope.  
%Radio properties of this  source is unlike IRS 7 which is 
%a radio-continuum-emitting and Brackett-line
%emitting source with a cometary tail  
%(Yusef-Zadeh, Morris and Ekers 1989; Rieke
%and Rieke 1989; Yusef-Zadeh \& Melia 1992). 
%A weak and compact radio continuum source 
%was detected from  the IRS 3's central star at a level  
%of 74 $\mu$Jy at 34 GHz  (Yusef-Zadeh et al. 2015). 

%The  cometary tail of ionized gas from IRS-7 points  directly away from the Galactic center.  

\section{Observation and Data Reduction}

ALMA  and
Karl G. Jansky Very Large Array
(VLA)\footnote{Karl G. Jansky Very Large Array (VLA) of the National Radio
Astronomy Observatory is a facility of the National
Science Foundation, operated under a cooperative agreement by Associated Universities, Inc.}
observations were carried out as part of a multi-wavelength observing campaign to monitor the
flux variability of Sgr~A*. 
Here we focus on
observations related to IRS 3 and IRS 7 within a few arcseconds  
of Sgr A*.
Observations were obtained  on  2016, July 12 and July 18.

% as part of directors' 
%discretionary
%time  to participate in  the observing campaign.

The ALMA 230 GHz data consisted of two spectral windows centered on
218.3 and 238.0 GHz, each 1.87 GHz wide.
Bandpass and delay calibration was based on J1924-2914.  Cross hand
gain calibration was based on Titan and Pallas which were assumed to
be unpolarized and subsequent calibration averaged the parallel hand
(XX and YY) data sets.
Initial amplitude and phase calibration was based on 1744-3116 with an
assumed flux density of 0.26 Jy at 234 GHz.
Phase self calibration followed by amplitude and phase calibration were 
applied. The amplitude self-calibration, however,  adds uncertainty to the
overall amplitude gain calibration.
The editing and calibration of the data was carried
out using OBIT (Cotton 2008) before all the spectral windows were averaged prior to constructing
final images.  
The July 12 data
with a spatial resolution of $0.36''\times0.25''$ are  presented here. 

We also present  radio data at 44 GHz and 15 GHz data taken with  the VLA in  its A configuration. 
The 44 GHz observations were 
carried out on 2011, July 8--9, and August
31--September 1. 
We combined all four days of observations resulting in about 25h of data, 
self-calibrated in phase and amplitude before final 
images were constructed. Details of these narrow band 
observations are given in Yusef-Zadeh et al. (2013). 
The 15 GHz observation,  which 
was part of a series of measurements at several frequencies between 1.4 and 44 GHz,  
was carried out on 2013, March 10. 
Data reduction and observing setup were identical to that described 
at 8 GHz in Table 1 of Yusef-Zadeh et al. (2016).

\section{Results}
\subsection{Radio and mm Emission from IRS 3}

Figure 1a,b show 226 GHz images of the central $10''\times10''$ 
of the Galactic center  
at  two different spatial resolutions
to reveal mm emission from  IRS 3 and  the newly extended structures. 
The peak mm emission from IRS 3 coincides with a compact point source at 44 GHz (Yusef-Zadeh et al. 2015).
Table 1 shows the comparison of radio and mm emission from IRS 3 
based on 
data taken on July 12, 2016. 
Columns of Table 1 give the frequency, 
the telescope and its  configuration, RA and Dec, the
angular distance from Sgr A* in increasing order, positional accuracy,
the spatial resolution, the peak intensity, the spectral index,  integrated intensities and  references.
The intensities  are estimated from background subtracted 
2D Gaussian fits to IRS 3. Table 1 also includes  data taken with the VLA in its A-array configuration on 2014,
March 9  and  2014, February 21 
at 34 and 44 GHz, respectively (Yusef-Zadeh et al.  2016).
The spectral index  $\alpha$, where
the flux density S$_\nu\propto\nu^{-\alpha}$,  between 44  and 226 GHz 
as well as  between 34 and 44 GHz give  values 1.85$\pm1.13$
and  1.32$\pm0.32$, respectively.  Using  the peak  flux density of IRS 3 at 350 GHz 
given by Moser et al. (2016), 
we find the spectral index is  $\alpha=1.17\pm0.33$ between 226 and 350 GHz. 
Radio emission from the photosphere of an evolved star has a typical spectral index close to 
$\alpha$=1.87 (Matthews, Reid and Menten 2015) whereas ionized mass-losing 
stellar winds  are characterized to have  $\alpha\sim0.6$ (Panagia \& Felli 1975). 
 It is clear that radio to  mm emission from IRS 3 is optically thick 
and its spectral index within errors 
is more consistent with the emission from the photosphere of an evolved  star than that of ionized stellar winds 
emanating from young mass-losing stars. Given the  large error in $\alpha$, 
we can not rule out the possibility that IRS 3 is a 
young mass-losing star. However, 
 spectral index measurements of a number of radio stars associated with young massive stars 
near Sgr A* do not show a steep optically thick  spectrum with high value of $\alpha$. 
In addition,  the spatially resolved structure of IRS 3 
 at 3.8$\mu$m combined with extended mm 
emission in its vicinity, as discussed below,  are detected only toward IRS 3 in the Galactic center. 
Thus, we assume that IRS 3 is an evolved AGB star.   

% based only on our  spectral index measurements. 

%IRS 3 is spatially 
%resolved in one direction with angular sizes  
%$\sim$21(168) and 69(552) mas(AU) with position angle (PAs) 123$^\circ$ and 
%169$^\circ$ (east of north) at 44 and 34 GHz, respectively. 

%Hoewever, our intensities may have been contaminated by the  mass-losing WR star IRS 3E 
%which is only 120 mas away from IRS 3. 
%Our radio and mm observations can not detect IRS 3E.

\subsection{Asymmetric  Shells and Fingers of Dust Emission}

ALMA images reveal extended emission from IRS 3  which includes 
four new substructures at mm wavelength.  The new mm substructures are faint but we detect them 
in both epoch of our mm observations. First, we detect mm emission from a shell of dust 
closest to the 
central star. Figure 1b shows this narrow layer of dust separated from the central peak, facing south towards Sgr 
A*.  The typical flux density of this layer of dust is 100 $\mu$Jy per $0.49''\times0.38''$ beam 
(PA$\sim-75^\circ$).  
Comparison between Figures 
1a,b and 1c shows that the innermost mm dust shell traces the southern edge of the dusty and cool AGB star at 3.8 
$\mu$m. The dark dashed lines closest to the central star outline the boundary of the dust shell at mm in Figures 
1a,b. The dashed white line outlines schematically the northern and southern  boundaries of the EW extended envelope in 3.8$\mu$m  (see Fig. 
1c). Second, a thick layer with a dearth of emission is sandwiched between  the inner circumstellar shell lying within 
1$''$ of the central star and an outer shell, about 2.5$''$ from the peak mm emission. The intensity in this 
dark layer ranges from 100 to --400 $\mu$Jy across the two mm  shells. 
We also note that 
the southern segment of IRS 7 in Figure 1a,b shows a similar dark feature. The outer dark dashed 
lines drawn 
on Figure 1 trace the outer shell of IRS 3.  The two shells and the dark layer of IRS 3 face Sgr A* with similar 
curvature suggesting that these shells are concentric and are associated with the central AGB star. 
We also note  
weak 15 GHz  continuum emission from the outer shell of IRS 3, as shown in Figure 1d. 
The mean intensity of the 15 GHz emission is  0.1 mJy per 0.2$''\times0.1''$ beam.    
Third, the outer mm shell is irregular in its appearance and reveals a 
number of ``fingers" of dust emission in Figure 1a,b. These fingers extend for about 2$''$ 
from IRS 3 towards  Sgr A* with  a position angle of $\sim120-160^\circ$.  

Figure 2a shows a 226 GHz grayscale close-up of 
the dark layer, the central mm peak emission from the photosphere of IRS 3, 
and the inner and outer circumstellar shells of IRS 3.  The peak emission appears to be extended in the NE-SW
direction is similar to that seen in the 3.8$\mu$m  with a PA$\sim65^\circ$ and angular size of $\sim3''$, as seen 
in  Figure  1c. The extended emission to the south of IRS 3 coincides with CS 
(5-4) emission and spots of compact SiO (6-4) emission
with radial velocities between 20 and 100 km s$^{-1}$ within 
a $6''\times4''$ region to the south of IRS 3 (Moser et al. 2016).  
The extinction map based on near-IR observations, as shown in Figure 2b,  also reveal a cloud  of extinction with 
an excess of $\sim$0.5 magnitudes with respect to its surroundings at 2$\mu$m (Sch\"odel et al. 2010), 
coincident with the  CS (5-4) emission. 
Figure 2c shows a close up view of IRS 3 at mm where 
contours of  226 GHz emission 
are superimposed on a 3.8$\mu$m grayscale image. This relatively low-resolution mm image 
shows that elongation of IRS 3 in the EW direction similar to that seen in Figure 1c. 
Figure 2d 
shows a schematic diagram of the new mm substructures, respectively. 

% Moser et al. (2016) have recently shown that the radial 
%velocities of CS (5-4) line emission between 
%delineated by the N arm of the mini-spiral, Sgr A*, IRS 3 and IRS 7 (see Fig. 6 of Moser et al. 2016). 

%We note that the most prominent fingers of dust emission at 225 GHz follows a ridge of CS 
%(5-4) emission extending for $\sim2''$ with the same position angle as those of the dust fingers 
%in the direction of Sgr A* with a position angle of $\sim130^\circ$. 
%The typical flux density of CS (5-4) line emission ranges between 1.5 and 2 Jy beam$^{-1}$ 
%\kms\, at  a resolution of 
%0.72$''\times0.57''$ (Moser et al. 2016).  The CS(5-4) radial velocity along the dust fingers ranges between 
%20 and  80 \kms.  Spots of SiO(6-4) line emission have  also been detected along this ridge (Moser et al. 2016). 
%Low resolution OH absorption study of the same region shows a peak toward the same region where
%the mm  and CS (5-4), SiO (6-5)  line emission have been detected (Karlsson et al. 2015).  
%The molecular counterpart to fingers and shells of dust emission 
%suggests that the source of molecular material with its finger-like appearance  
%arises mainly from the envelope of IRS 3 that is infalling toward Sgr A*. 

%The  diffuse mm dust shell 
%follows the sharp  southern boundary of IRS 3 as detected in mid-IR. 

\subsection{Radio Dark Dusty Stars}

%In the case of IRS 7, 
%the outer ionized feature coincides with  a bow shock structure (
%Yusef-Zadeh \& Melia 1992). Radio continuum images also show new structures at the position of IRS 3. 
%In addition, Figure 1d shows a 15 GHz image 
%revealing  a dark feature  at the position of 
%IRS 3 within the dusty shell at mid-IR and inner shell at mm. This dark feature is outlined
%by the dark dashed lines in Figure 1d. 
%The drop in the 15 GHz background intensity at the position of the dusty shell is 
%a factor of 2.5.  The dark radio feature coincident with IRS 3 is the  imprint of dusty envelope 
%There are also deeper dark features near the N arm of the mini-spiral  and IRS 7. 

%A second dark feature lies 
%to the NE where IRS 7 is located. 
%This figure shows the emission from the photosphere of IRS 3 (see Table 1) and a cometary structure associated
%with  IRS 7. 
%For comparison Figure 3b shows a 3.8$\mu$m image with identical size to that of Figure 3a. 

Radio continuum images show 
dark features toward the envelopes of IRS 3 and IRS 7, 
reminiscent of radio dark clouds (RDCs) 
that appear  in radio continuum images of 
Galactic center sources (Yusef-Zadeh 2012). Radio dark clouds  in continuum images with high dynamic images 
provide imprints of 
molecular and dust clouds that are embedded in a bath of 
ionizing radiation. 
In regions where  thermal radio continuum emission is depressed along the line of sight 
through a molecular cloud, the cloud  appears as a dark feature in radio 
continuum images.  This depression can also be produced by swept up gas by an outflow 
within an ionized medium. 

%Because these dark features are not conspicuous,
Figure 3a-d present 
the dark features associated with  IRS 3 at 44 GHz and 3.8$\mu$m. 
The locations of three  dark features  are drawn as  ellipses. 
The one to the east coincides with  the region surrounding the northern  arm of the mini-spiral. 
The dark features to the southwest and northwest  lie in the vicinity of the IRS 13N complex (Muzic et al. 2008) 
and IRS 3, respectively.  The feature close to IRS 3 traces 
roughly the  shells  noted in Figures 1 and 2. 

A close-up  and high resolution view of the region around IRS 3 and IRS 7  is 
presented in Figure 4a,b at 44 GHz and 3.8$\mu$m.  A faint radio source which 
coincides with  the central star of IRS 3 (see Table 1) is labeled. 
We note an extended and elongated dark feature to the southwest of 
IRS 3. This dark feature, drawn as white broken lines,  
covers the inner shell and the dark layer detected at mm, as described above. 
We also note another dark feature
in the immediate vicinity of IRS 7.  Both these dark features extend to the outer envelope 
of the dusty stars IRS 3 and IRS 7 and terminate  where strong and weak   ionized gas is detected, respectively. 

%IRS 1W coincides with clumps of dust emission suggesting that star formation activity is 
%taking place (Tsuboi et al. 2016).  
%Most recently, ALMA observations have detected CS(5-4) line emission 
%extending several  arcseconds to  the north of Sgr A* surrounding IRS 3 and IRS 7. 

Dark features  in radio continuum images 
could result from 
interferometric errors due to the incomplete sampling of the {\it {uv}} plane  (Yusef-Zadeh 
2012).  However, we  detected these dark features in several high frequency images using the VLA in different 
array configurations. In addition,  
the dark features correlate with a reservoir of 
molecular gas  in the region between IRS 3, IRS 7 and Sgr A* (Moser et al.  2016). 
The expected correlation of molecular line emission surrounding IRS 3 
and radio dark clouds is  consistent with that noted toward radio dark clouds  and molecular 
clouds (Yusef-Zadeh 2012). In addition, dark clouds  with 
excess near-IR extinction (Sch\"odel et al. 2010). 
The CS (5-4) line emission and an extinction excess in near-IR 
(Moser et al. 2016; Sch\"odel et al. 2009) 
coincide with radio  dark dusty star IRS 3 and its extension to the south east. 
The third dark feature to the NE, however, shows no molecular or extinction counterparts. 
This feature and the one surrounding the  inner 2$''$ of Sgr A* coincide with 
dust cavities. 

%  and are likely produced by an outflow, discussed elsewhere. 

%an outflow within the ionized medium of the Galactic center.  
%The lack of any molecular line emission to the west of the N arm of the minispiral is consistent with this 
%suggestion. 

%IRS 13N is a site where a number of YSO's have been detected 
%(Muzic  et al. 2008). 

%This is because, the background nonthermal continuum emission becomes significant and dark 
%features disappear.

%The existing radio continuum data, however, 

%are not adequate to establish the identification of radio dark clouds (RDCs).  This is because radio dark clouds 
%can not be distinguished from interferometric artifacts, if the {\it {uv}} plane is not sampled uniformly. These 
%dark clouds which should be edge-brightened by the external radiation field should be detected in high dynamic 
%range radio continuum images.

\section{Discussion}

\subsection{Tidal Distortion of  the Circumstellar Shells of IRS 3} 

%The origin of these morphological features is not clear but is likely to be associated 
%with tidal disruption of past shells associated with IRS 3. Here we study  the elongation 
%of IRS 3 as seen in 3.8$\mu$m images and  sugges that the elongation is consistent with tidal 
%disruption of the envelope of IRS 3 as it orbits Sgr A*. 

%This highly excited line emission has critical density $\sim3\times10^6$ cm$^{-3}$.  

\newcommand{\cs}{\mathrm{CS}}
\newcommand{\nh}{n_\mathrm{H}}

The most interesting result of this study  is 
the discovery of asymmetric mm dust shells  centered on IRS 3 
and facing Sgr A*.
The shells exhibit a 
wavy pattern  with wavelength   $\sim0.25''-0.5''$
along the shell. In addition, the outermost shell of IRS 3 is
broken up into fingers of mm emission,    pointing  
toward Sgr A*.  The fingers of dust emission appear to coincide with a region where 
CS (5-4)  emission is concentrated (see Fig. 6 of Moser et al. 2016). 
We first  estimate the mass of gas associated with the CS(5-4) emission, and show that it could arise from the 
envelope of IRS 3.  The emission extends over a $\sim6\arcsec\times4\arcsec$ region with velocities between about 
40 and 80 \kms (Moser et al. 2016). 
We used the online 
version of \textsc{radex}\footnote{http://home.strw.leidenuniv.nl/~moldata/radex.html} (Van der Tak et al.\ 2007) to 
estimate the density and column of the emitting molecular gas.  Adopting a kinetic temperature of 300\,K, consistent 
with the proximity to hot stars in the central parsec, and a 40 \kms\,  FHWM,  
the observed radiation temperature can be 
produced by a medium in which the product of hydrogen number density and line-of-sight CS column $\nh N_\cs \approx 
1\times 10^{19}$\,cm$^{-5}$ as long as $\nh\,\la 3\times 10^5$\,cm$^{-3}$.  
Under these conditions the line is 
sub-thermally excited and optically thin. 
We assume that the CS abundance 
relative to H to be $x_\cs = 5\times10^{-7}$,
consistent with observations of carbon stars (Woods et al. 2003) 
and assume that 
the depth of the source along the line of sight, $L$, is 0.2\,pc, consistent with the angle subtended by the CS 
emission on the sky.  Then noting that $\nh\,N_\cs = x_\cs\,\nh^2\,L$, we find $\nh \approx 
6\times10^3$\,cm$^{-3}$.  This implies an uncomfortably large envelope mass $M_\mathrm{env}\approx 1.4\msol$
unless the IRS 3 envelope is clumpy giving 
$M_\mathrm{env}=0.3 \msol$ with most mass residing in clumps with 5\% 
filling factor (implying $\nh\approx3\times10^4$\,cm$^{-3}$). This 
 would produce the observed CS 5-4 line emission.  
Although this envelope contains 1/5 the original number of CS molecules, each lies in a clump with five times the 
original density: the rate of collisional excitation of each CS molecule, and hence the rate of J=5-4 photon 
emission per CS molecule, is increased fivefold. Individual clumps have $\sim0.8$ magnitudes of extinction at 
$2\mu$m. This can be reconciled with the observed average extinction (see Fig, 2b) if the beam filling fraction 
is $\sim$0.3. The volume and area filling fractions imply that the envelope consists of $\sim16$ clumps with size 
$\sim0.014$ pc, or $0.7''$. Adopting a dust temperature of 100 K in the envelope implies an average thermal 
continuum $\sim90\, \mu$Jy per $0.39''\times0.38''$ beam at 226 GHz, similar to the mm enhancement coincident 
with the CS emission and IR extinction, are broadly consistent with the structure seen in the mm emission.

%, with a radiation temperature $\sim0.5\,$ K (see Fig.\ 7 of Moser et al. 2016).  

% so that almost every molecule that is collisionally excited to the J=5 
%rotational level spontaneously decays to J=4, with the photon escaping the source.  

%typical of the envelopes of carbon stars (reference), 

%However, AGB envelopes are clumpy, and s
%No IR lines have been detected from the envelope of IRS 3, but 

Infrared $^{13}$CO ($v=1-0$) and H$^+_3$ absorption lines have also been detected toward IRS 3 (Goto et al. 2014). 
One broad velocity component is at 60 km s$^{-1}$ ranging between 51 and 85 km s$^{-1}$. This velocity feature 
has not been detected toward IRS 1W a few arcseconds away from   Sgr A*. 
The physical characteristics of H$^+_3$ suggest density and temperature that are consistent with above 
estimates from CS measurements. Goto et al. (2014) interpret 
the 60 \kms feature arising  from the inner pc but  associated with 
the circumnuclear molecular ring. Given that the  infrared absorption lines have similar characteristics to the 
CS line emission discussed above, it is possible that absorption features 
 come from the envelope of IRS 3. Based on photospheric modeling, Pott et al. (2008) 
estimate $\dot{M}\sim 6\times10^{-5}$\,\msol\,yr$^{-1}$. The time scale to eject 0.3$\msol$ 
is then $\sim5\,000$\,yr, 
implying a terminal wind speed $v\sim 20$\,\kms\, to yield an envelope radius $r\sim0.1$\,pc, consistent with the extent 
of the CS emission.  The envelope may, of course, extend to greater radii but not be visible in CS if molecules are 
dissociated by the strong external FUV field at the Galactic center.  Alternatively, the envelope may be truncated 
if significant mass loss from IRS 3 has only been taking place over the past 5\,000 yrs, or it could be tidally 
stripped.

Tidal effects may be responsible for the  distortion of the envelope as IRS 3 orbiting  Sgr A*.
Alternatively, the asymmetry may be  due to an asymmetric outflow 
from IRS 3 as numerous post-AGB stars reveal this structure (Lykou et al. 2015; Richards et al. 2011).  
We can not  rule out this possibility but 
given the elongation of IRS 3 along its proper motion and the presence of 
extended molecular and dust emission distributed to the south 
of IRS 3, we consider 
 that the  elongation of IRS 3 at  3.8$\mu$m is  due to tidal distortion, as described below.

A simple criterion for tidal extension is 
that the expansion time scale is of order of the orbital time scale around Sgr A*:

$$r/v \approx (GM/R^3)^{-0.5} $$

where $R\sim0.8$ pc is the distance from Sgr A* 
 and $r$ is the shell radius.
To explore the envelope's distortion we model it 
as a set of fluid elements that are launched radially outwards from IRS 
3 as it orbits Sgr A* and subsequently follow their ballistic 
trajectories in the gravitational field of Sgr A*.  In other words, each 
fluid element follows an independent Keplerian orbit consistent with its 
launch position and velocity.  This ballistic approximation is 
reasonable as the outflow is supersonic and so the dynamical effect of 
pressure is negligible until the fluid elements start to intersect $\sim 
1/2$ an orbital period after launch.

%that the change in Sgr A*'s gravitational potential across the distorted shell is of order the kinetic energy of the 
%wind material, ie $$1/2 v^2 \approx GM/R^2 \times r $$
%Indeed, the potential tidal distortion of the envelope by Sgr A* places 
%constraints on the location of IRS 3 along the line of sight: IRS 3 must 
%be sufficiently distant from Sgr A* that it can retain an envelope on 
%$\sim0.1$\,pc scales.  

The orbit of IRS 3 is not completely determined a-priori because its line-of-sight velocity and location within the 
inner parsec along the line of sight are unknown, nor has acceleration in its proper motion across the sky been 
observed.  However, the projected location of IRS 3 relative to Sgr A* and its proper motion are both well 
determined (Sch\"odel et al.\ 2009).  We adopt $+50$\, \kms\, 
as the line-of-sight velocity, representative of the 
mean velocity of the CS emission that we assume arises from its envelope.  
Then we vary the line-of sight distance 
$z$ relative to the distance to Sgr A*.  A particular choice of $z$ means that the instantaneous velocity and 
displacement of IRS 3 relative to Sgr A* are specified, sufficient to specify 
its  orbit.   We adopt 
$4\times10^6\msol$ and 8\,kpc as the mass and distance of Sgr A*, and RA and Dec offsets of IRS 3 from Sgr A* as 
 $ \Delta RA = -2\dasec341 \pm 0\dasec009$, $\Delta Dec = 3\dasec848 +- 0\dasec016 $ with velocities $v_\mathrm{RA} = 
179.1 \pm 5.6$ \kms, $v_\mathrm{Dec} = 31.2 \pm 4.0$\,\kms (Sch\"odel et al.\ 2009).  We adopt a z-axis directed 
away from the observer, with $v_z = 50$\,\kms, and $z=0$ at the distance of Sgr A*, so e.g.\, $z=+0.1$\, pc lies 
0.1\,pc beyond Sgr A*.

We compute the ballistic evolution of fluid elements ejected from IRS 3,  2\,500, 5\,000, 7\,500\, yr  ago, and plot 
their position on the the sky relative to Sgr A* as black, blue , and red points in Figure 5, 
for different choices 
of $z$.  The blue shell should correspond to the extent of the CS envelope, which has an estimated flow time 
$\sim5\,000$\, yrs.  If $z\la -0.8$\,pc, or $z\ga 0.6$\,pc, the tidal distortion of the shell is too small, whereas 
if $-0.6\,\mathrm{pc}\la z\la0.4$\,pc the tidal distortion is too severe.  We therefore conclude that if the CS 
emission arises from the envelope of IRS 3 and  IRS 3 
currently lies either $\sim0.7$\,pc in front of Sgr A*, or 
$\sim0.5$\,pc behind it.

According to Viehmann et al. (2005), the 3.8$\mu$m  isophotes of IRS 3 have major axis orthogonal 
to the direction towards  
Sgr A*, and  have a bow shock morphology perhaps caused by winds from the cluster of massive stars orbiting 
Sgr A* or an outflow from Sgr A*. However, this morphology is very dissimilar to the head-tail ionized structure 
associated with the bow shock source IRS 7 (Yusef-Zadeh, Morris and Ekers 1989; Rieke
and Rieke 1989; Yusef-Zadeh \& Melia 1992). 
 There is no evidence of an envelope of ionized gas surrounding the 
innermost mm shell of IRS 3. There is weak radio continuum emission from the outer distorted shell (see Fig. 2b). 
Thus, it seems unlikely that the the NE-SW  structure of IRS 3 at 3.8$\mu$m is produced by external winds. Instead, 
we suggest that the outer shell asymmetry is produced by tidal effects tending to stretch the envelope along the 
orbit of IRS 3. It is also possible that the dust shells to the south of IRS 7 could be generated 
by the tidal tails
from an earlier episode  of mass loss from IRS 3 (see the z=+0.4 panel in Fig. 5).

In summary, mm and radio images of the inner 10$''$ of Sgr A* reveal stellar and circumstellar 
emission from IRS 3.  The envelope of this AGB star is argued to be distorted and 
 disrupted by the tidal 
force of Sgr A*. 
We also showed that dusty envelope of stars could have their imprints on radio 
continuum images. These dark radio stars are produced the same way that radio dark clouds 
are originated by being embedded within an ionized medium.

%This interaction could be responsible for supplying  cold and dense gas 
%to Sgr A*. 

%closest to Sgr A*, destroying it before it stretched it in the radial direction.  The outer 
%shell may be %recognized as the shell that was removed from the envelope of the star and the 
%dark feature at mm %may support this. Alternatively, the dark layer may have been generated by 
%the presence of molecular gas with %low dust-to-gas ratio.

\acknowledgments
This work is partially supported by the grant
AST-1517246 from the NSF.
RS acknowledges funding from the European Research Council under
the European Union's Seventh Framework Program (FP7/2007-2013) 
/ ERC grant agreement  [614922].
MW acknowledges the hospitality of the Department 
of Physics \& Astronomy at New Mexico State University, where part of this work was carried out.
This paper makes use of the following ALMA data:
2015.A.00021.S (Principal Investigator Gunther Witzel)
ALMA is a partnership of ESO (representing its member states), NSF (USA) and NINS
(Japan), together with NRC (Canada) and NSC and ASIAA (Taiwan), in cooperation with the
Republic of Chile. The Joint ALMA Observatory is operated by ESO, AUI/NRAO and NAOJ.
We thank S. Gillessen for providing us with a 3.8$\mu$m image.

%\newcommand\refitem{\bibitem[]{}}

%\vfill\eject

\begin{deluxetable}{ccccccccccc}
%\rotate
\tabletypesize{\scriptsize}
\tablecaption{Parameters of 2D Gaussian fits to IRS 3}
\tablecolumns{11}
\tablewidth{0pt}
\setlength{\tabcolsep}{0.05in}
\tablehead{
\colhead{Freq.} & \colhead{Telescope} & \colhead{RA}                       &
    \colhead{Dec}     & 
    \colhead{Sgr A*} & \colhead{Pos.}     & \colhead{$\theta_{a}\times\theta_{b}$(PA)} & 
    \colhead{Peak}      & \colhead{Spectral} & \colhead{Integrated} & \colhead{Refs} \\

\colhead{(GHz)} & \colhead{(Config)}  & \colhead{(J2000)}                  &
    \colhead{(J2000)} &
    \colhead{Offset} & \colhead{Accuracy} & \colhead{}\sl{mas}$\times$\sl{mas}(\sl{deg})                                  &
    \colhead{Intensity} & \colhead{Index}    & \colhead{Flux}       & \colhead{}     \\

\colhead{}     & \colhead{}           & \colhead{($17^{\rm h}45^{\rm m}$)} & 
    \colhead{($-29^{\circ}00^{\prime}$)} & 
    \colhead{(\sl{arcsec})} & \colhead{(\sl{mas})} & \colhead{} &
    \colhead{\sl{(mJy beam$^{-1}$)}} & \colhead{($\alpha$)} &  \colhead{\sl{(mJy)}} & \colhead{}
}
\startdata
44.5  & VLA(B)  & 39.8614 & 24.2147 & 4.50 & 50.24 & 223$\times$127(3.8)
    & 0.109 $ \pm $ 0.052 & --  &  0.121 $ \pm $ 0.096 & \\
226  & ALMA  & 39.8644 & 24.2573 & 4.44 & 101.29 & 442$\times$341(-70.8)
    & 0.841$\pm$0.021 & 1.32$\pm0.32$\tablenotemark{a} & 1.215$\pm$0.400 &  \\
350  & ALMA &          &        &      &        & 490$\times410$                  
    &  1.4$\pm0.2$        &  1.17$\pm0.33$\tablenotemark{b}    &  1.22$\pm0.38$    &  c\\
\hline
34.5  & VLA(A) & 39.8619 & 24.2522 & 4.47 & 10.93 & 85$\times$42(-5.3)
      & 0.081 $ \pm $ 0.012 & --                & 0.108 $ \pm $ 0.025 & \\
44.5  & VLA(A) & 39.8624 & 24.2520 & 4.46 & 7.55 &  74$\times$34(-4.0)
      & 0.132 $ \pm $ 0.034 & 1.85$\pm1.13$  & 0.111 $ \pm $ 0.052 & \\
\enddata
\vspace{-10pt}
\tablenotetext{a}{$\alpha$ between 44.5 and 226 GHz}
\tablenotetext{b}{$\alpha$ between 226 and 350 GHz}
\tablenotetext{c}{Moser et al. 2016}
\end{deluxetable}

%grayscale flux -0.1 to 300 mJy beam$^{-1}$). 
\begin{figure} 
\center
\includegraphics[scale=0.25,angle=0]{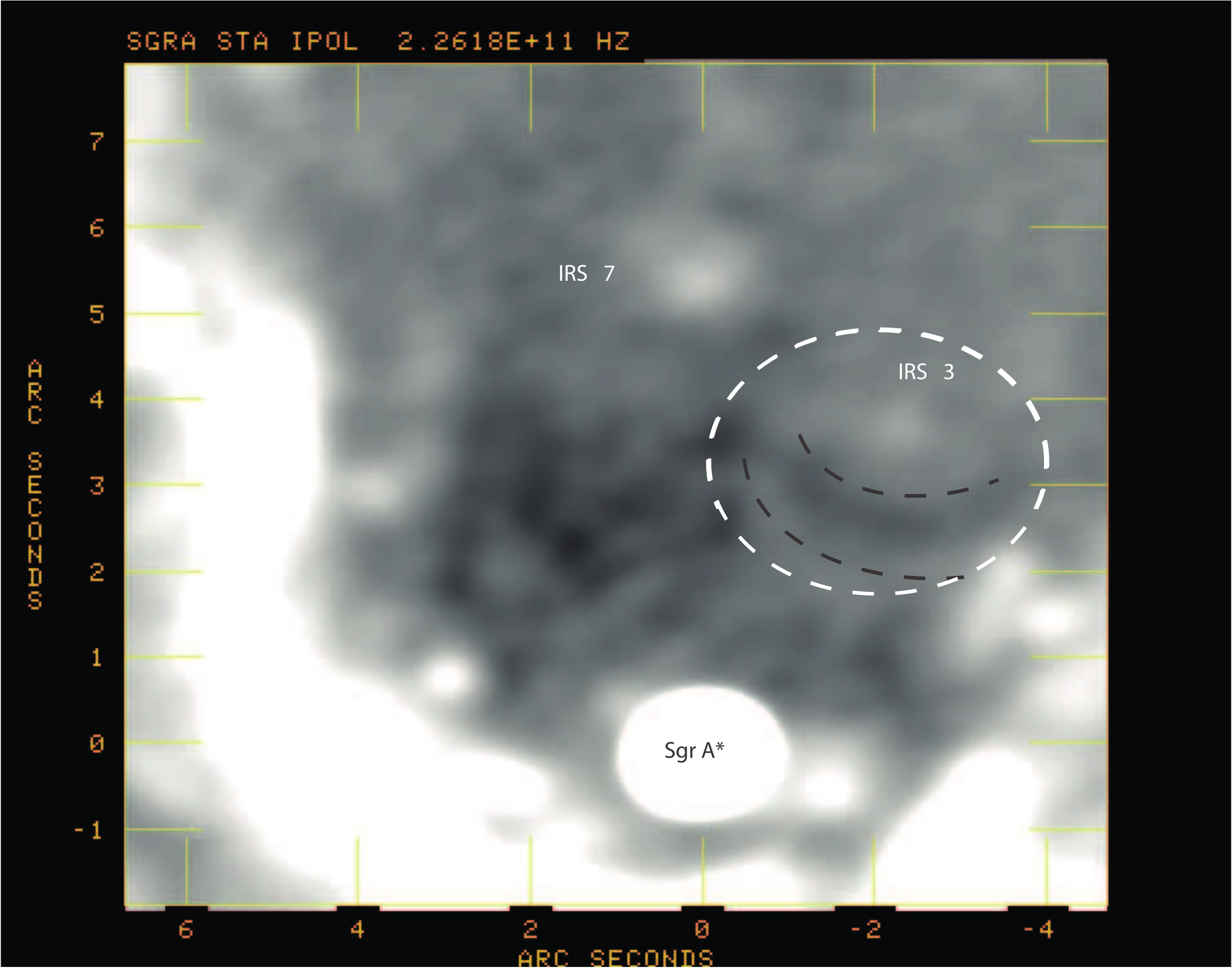}\\ 
\includegraphics[scale=0.25,angle=0]{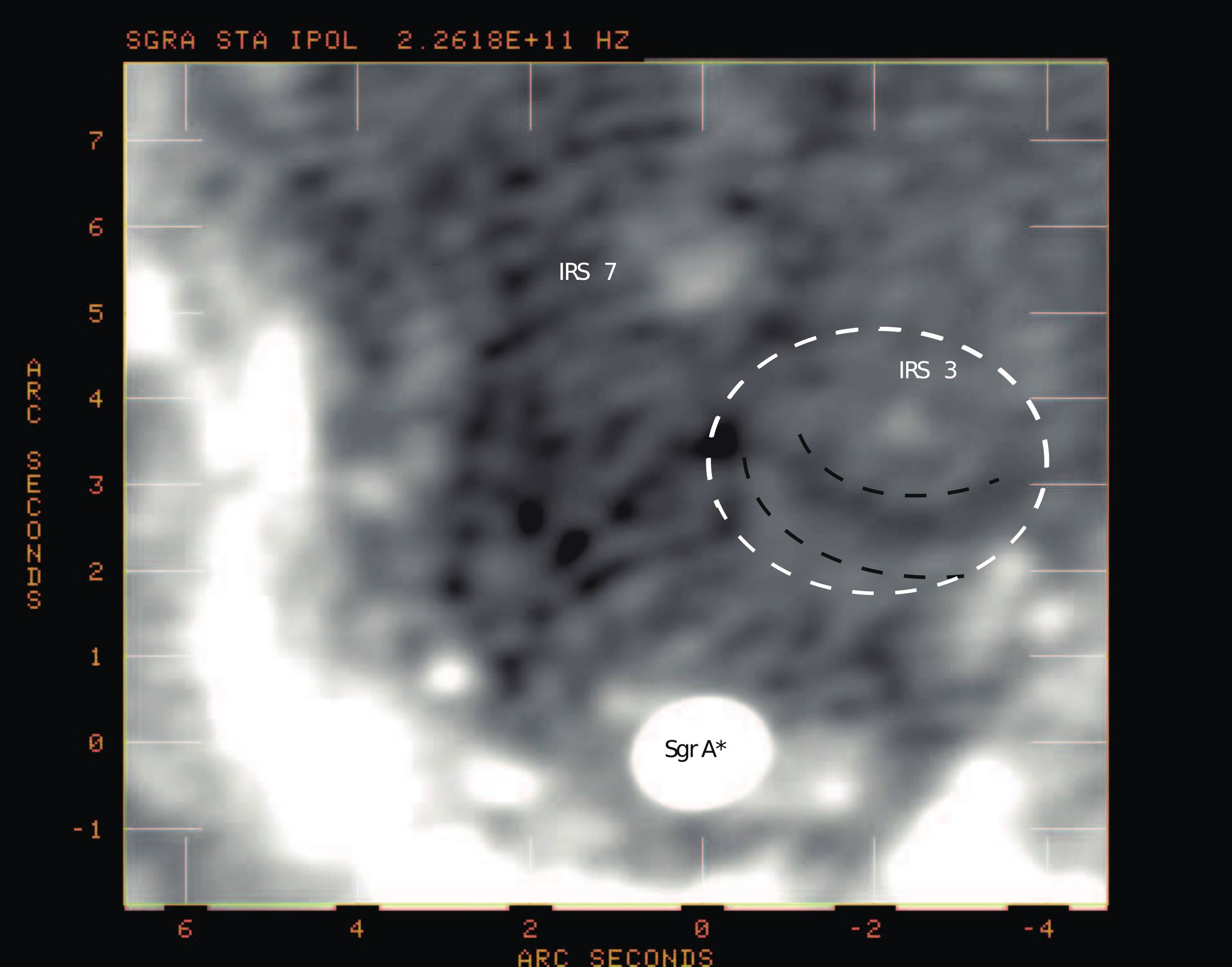}\\
\includegraphics[scale=0.25,angle=0]{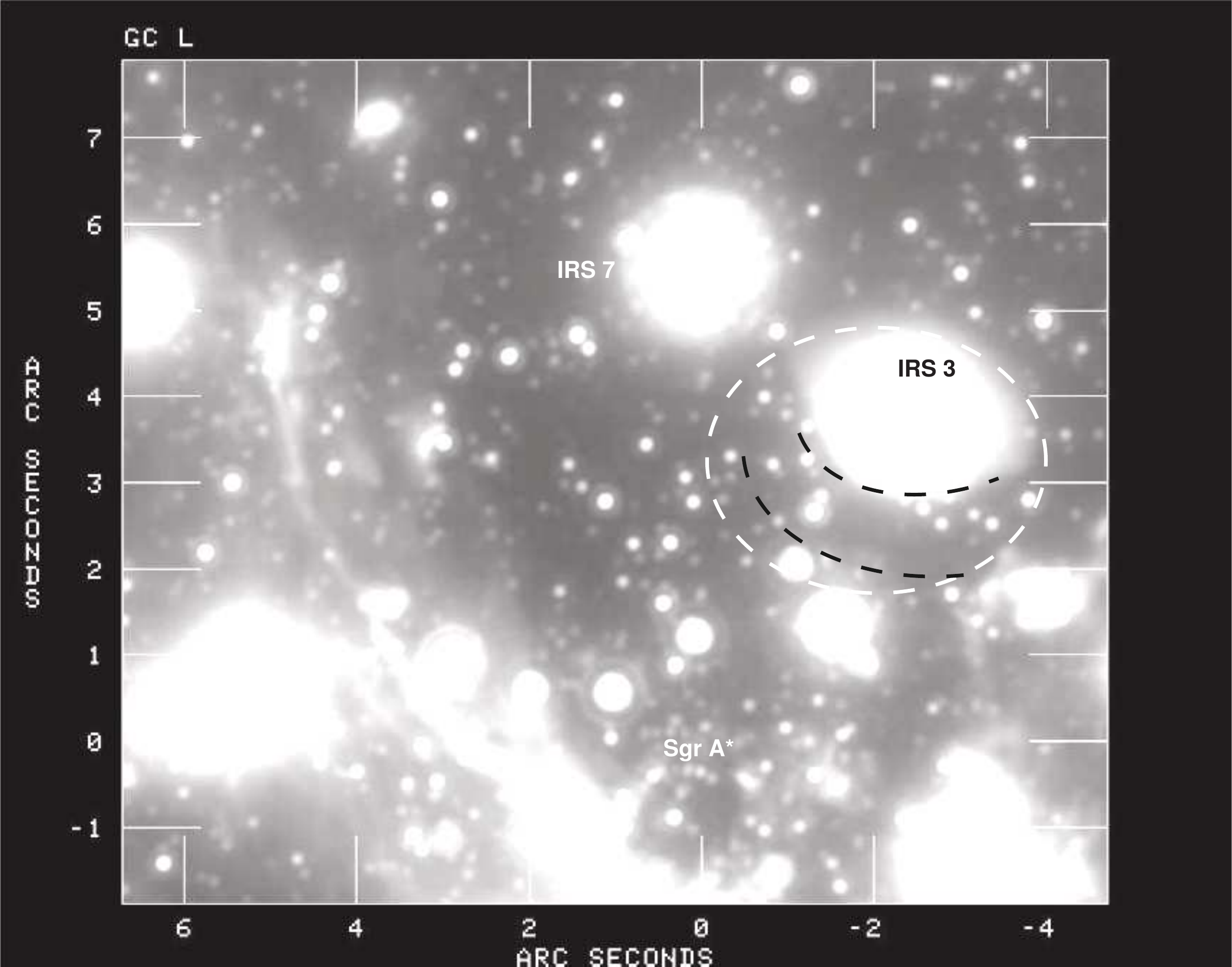}\\ 
\includegraphics[scale=0.25,angle=0]{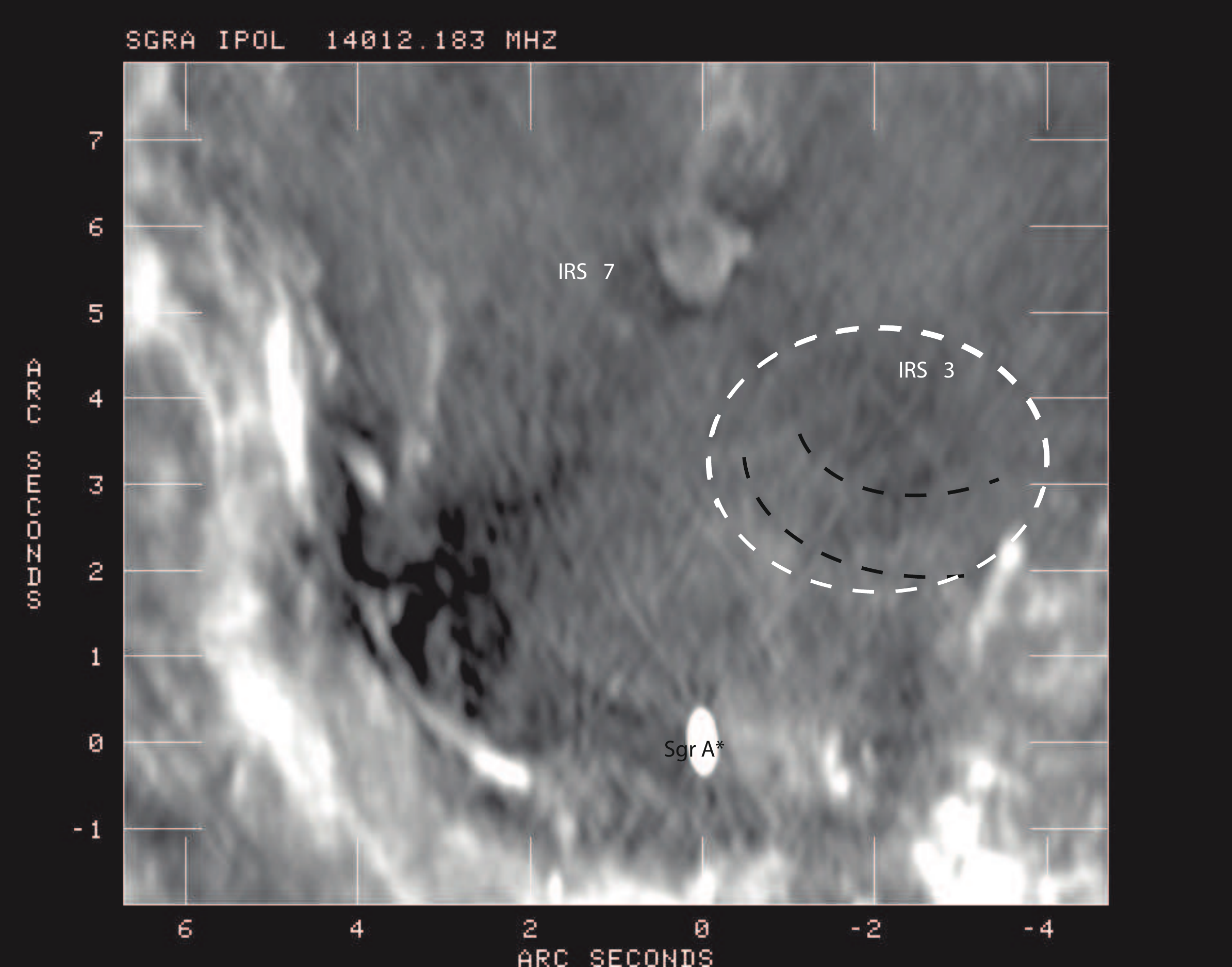} 
\caption{
{\it (a) Top} A grayscale mm image of the inner 10$''\times10''$ 
of the Galactic center at 226 GHz with a
resolution of $0.49''\times0.38''$ (PA=$-75^{\circ}$) taken with ALMA.  
{\it (b) Top Middle} Similar to (a) except with a resolution of
 $0.44''\times0.34''$ (PA=$-70^{\circ}$).  
{\it (c) Bottom Middle} 
Similar region to that of (a) at 3.8$\mu$m taken with  the VLT (Gillessen, priv. comm.)
{\it (d) Bottom} 
A radio image of the same region shown in (a) at 15 GHz
at a resolution of $0.24''\times0.11''$ (PA=$3.8^{\circ}$)
taken with  the VLA on 10 March, 2014.}
\end{figure}

\begin{figure} \center 
\includegraphics[scale=.4,angle=0]{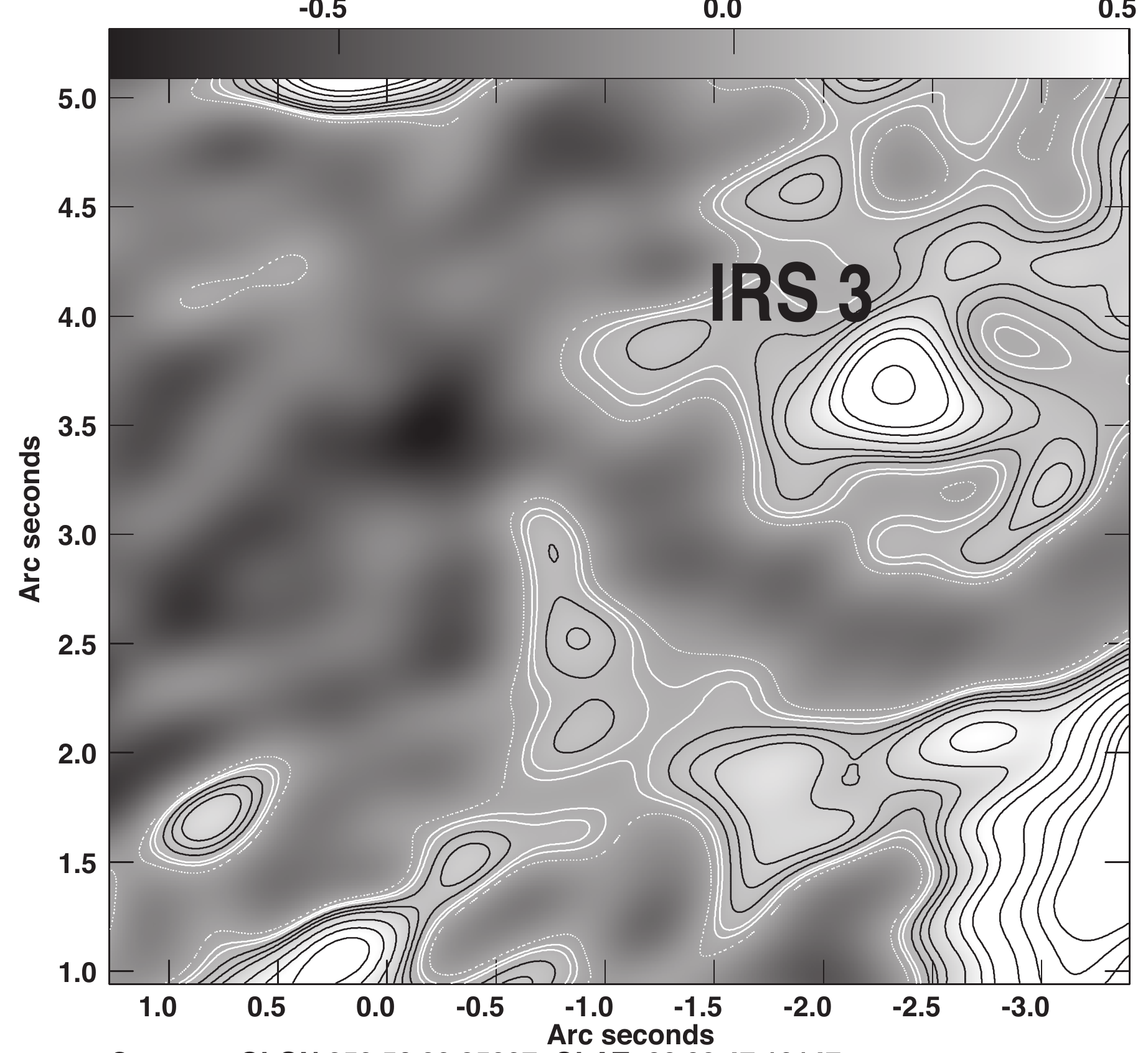} 
\includegraphics[scale=.4,angle=0]{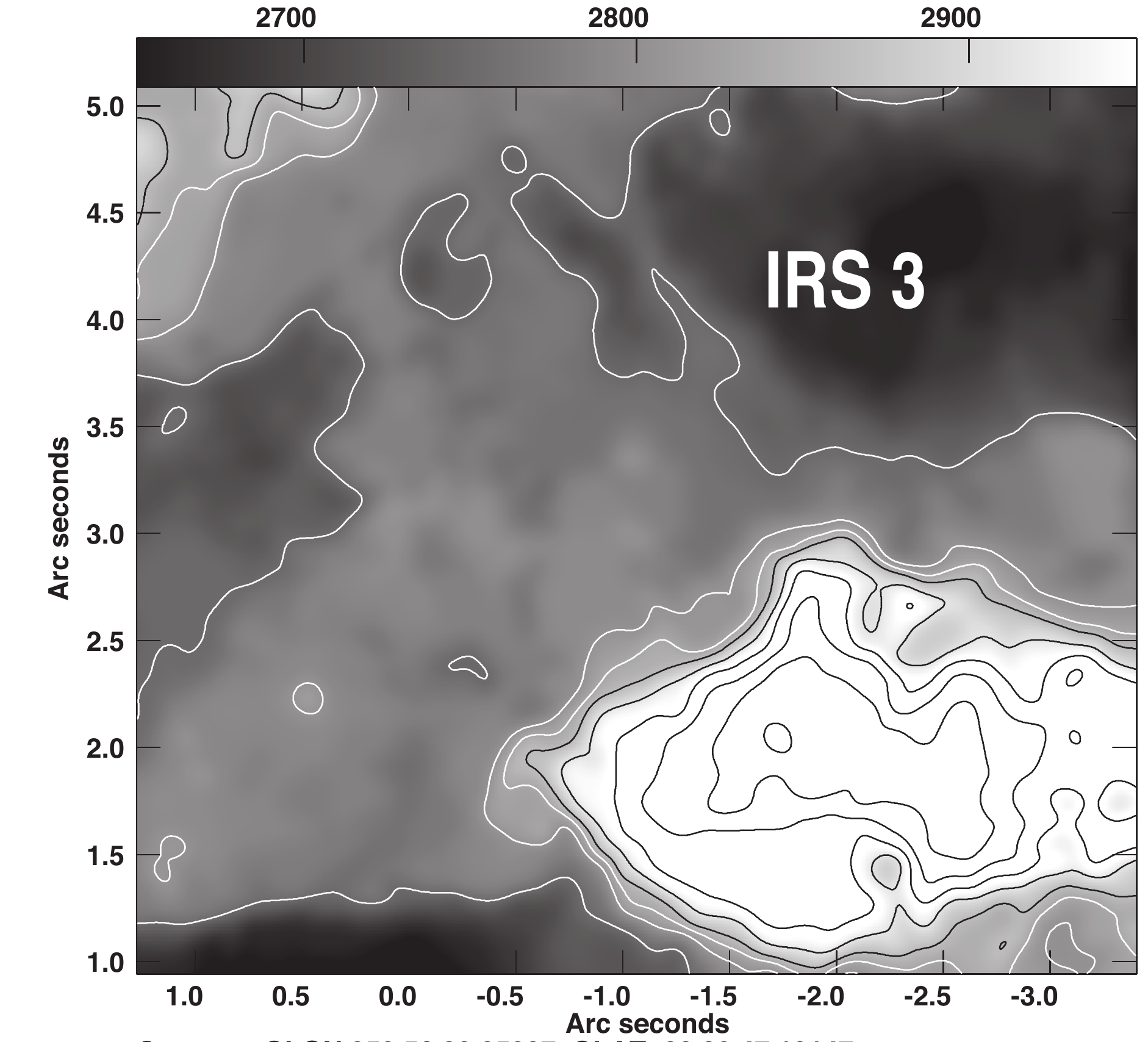} 
\includegraphics[scale=.4,angle=0]{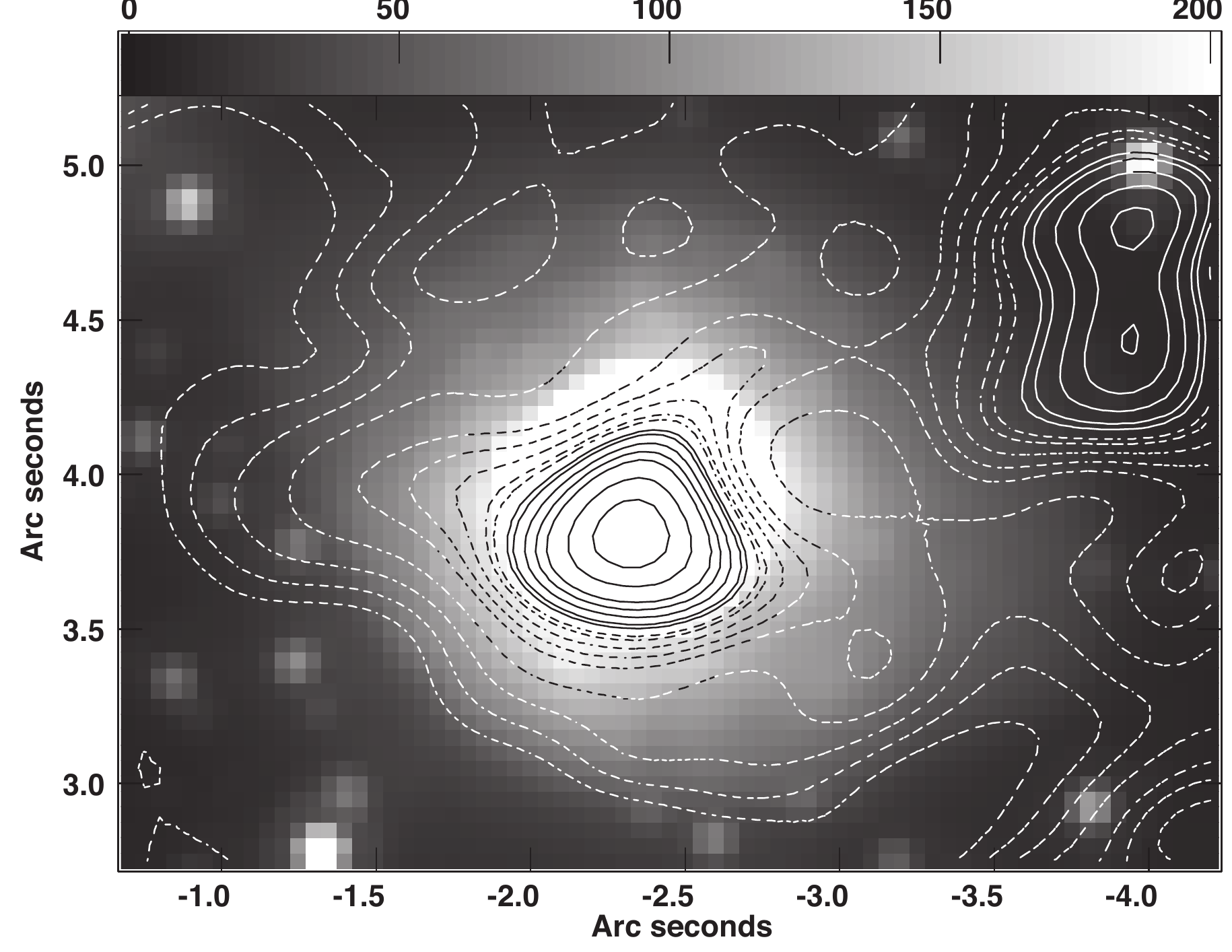} 
\includegraphics[scale=.2,angle=0]{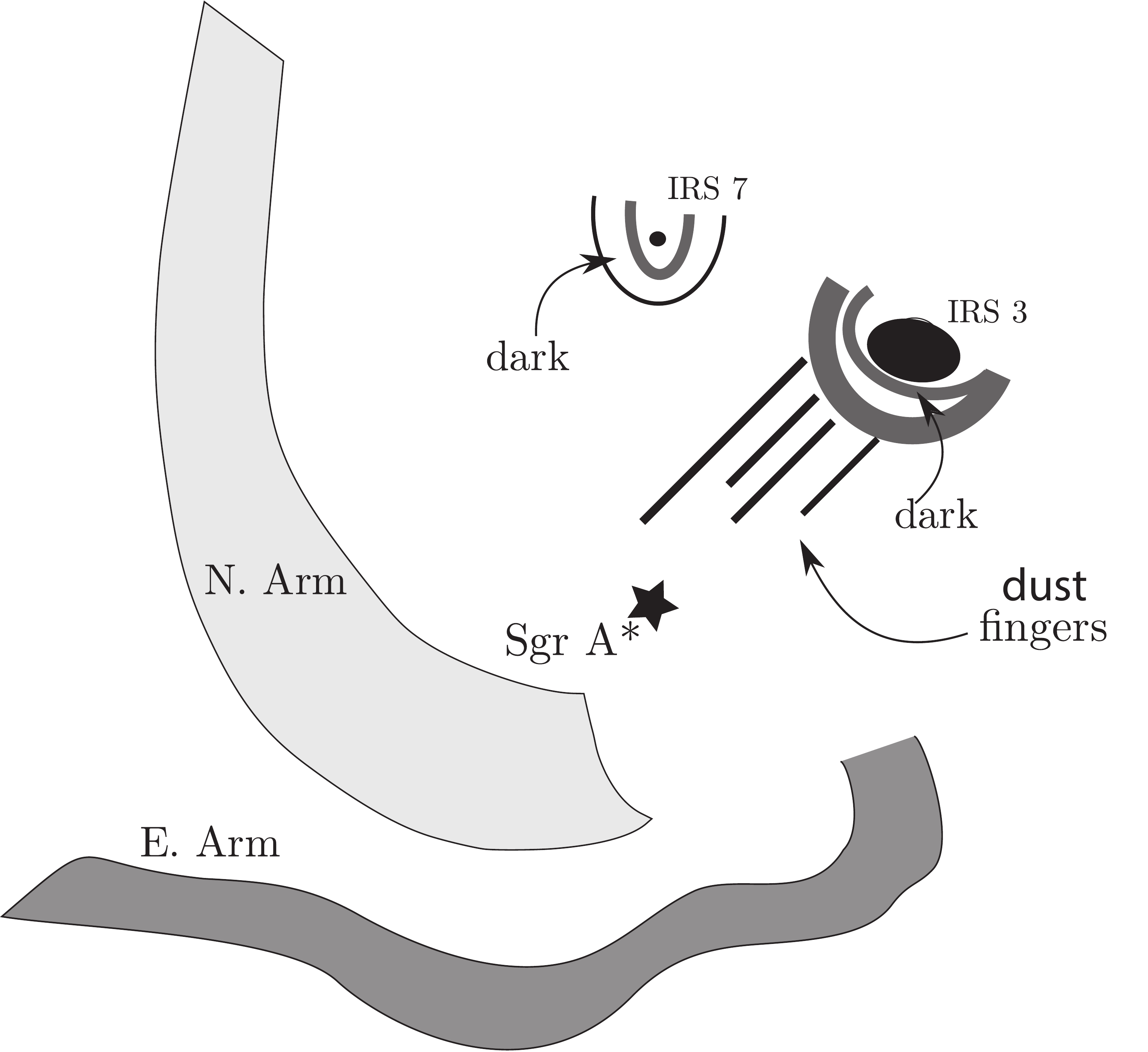} 
\caption{ 
{\it (a) Top Left} 
Contours of 226 GHz emission from the inner 2.5$''\times3.5''$ of IRS 3 are set 
at (-0.5, 0.5, 1, 2, 3, 4, 6, 8, 10, 15, 20, 25, \& 30)$\times55.5$ 
$\mu$Jy beam$^{-1}$ with a resolution  of 
$0.49''\times0.38''$ (PA=$-74^{\circ}$) with grayscale range between -0.79 and 0.5 mJy beam$^{-1}$.  
{\it (b) Top Right} Similar to (a) 
except that it is an extinction map 
with contours set at 2.5, 2.55, 2.57, 2.6, 2.65, 2.7, 2.75, 2.8,
2.85, 2.9, 2.95, 3, 3.025, 3.050) $\times1.1$ magnitudes at K band  (Sch\"odel et al. (2009).
The grayscale range is between 2.650 and 2.950 magnitudes.
{\it (c) Bottom Left} 
Contours of 226 GHz emission with a resolution of 0.59$''\times0.47''$ (PA$\sim-82.5^\circ$) 
are superimposed on a 3.8$\mu$m image of IRS 3 with levels set at 
at (-12, -10, -8, -6, -4, -3, -2, -1, -0.5, 0.5, 1, 2, 3, 4, 6, 8, 10, 12)$\times55.5$ mJy beam$^{-1}$.  
Given that this image is weighted to show extended features, 
emitting features are depressed by  the negative bowls due to lack of short spacings. 
{\it (d) Bottom Right} 
A schematic diagram shows the new mm sub-structures between IRS 3 and Sgr A*. 
}
\end{figure}

\begin{figure} 
\center 
\includegraphics[scale=.3,angle=0]{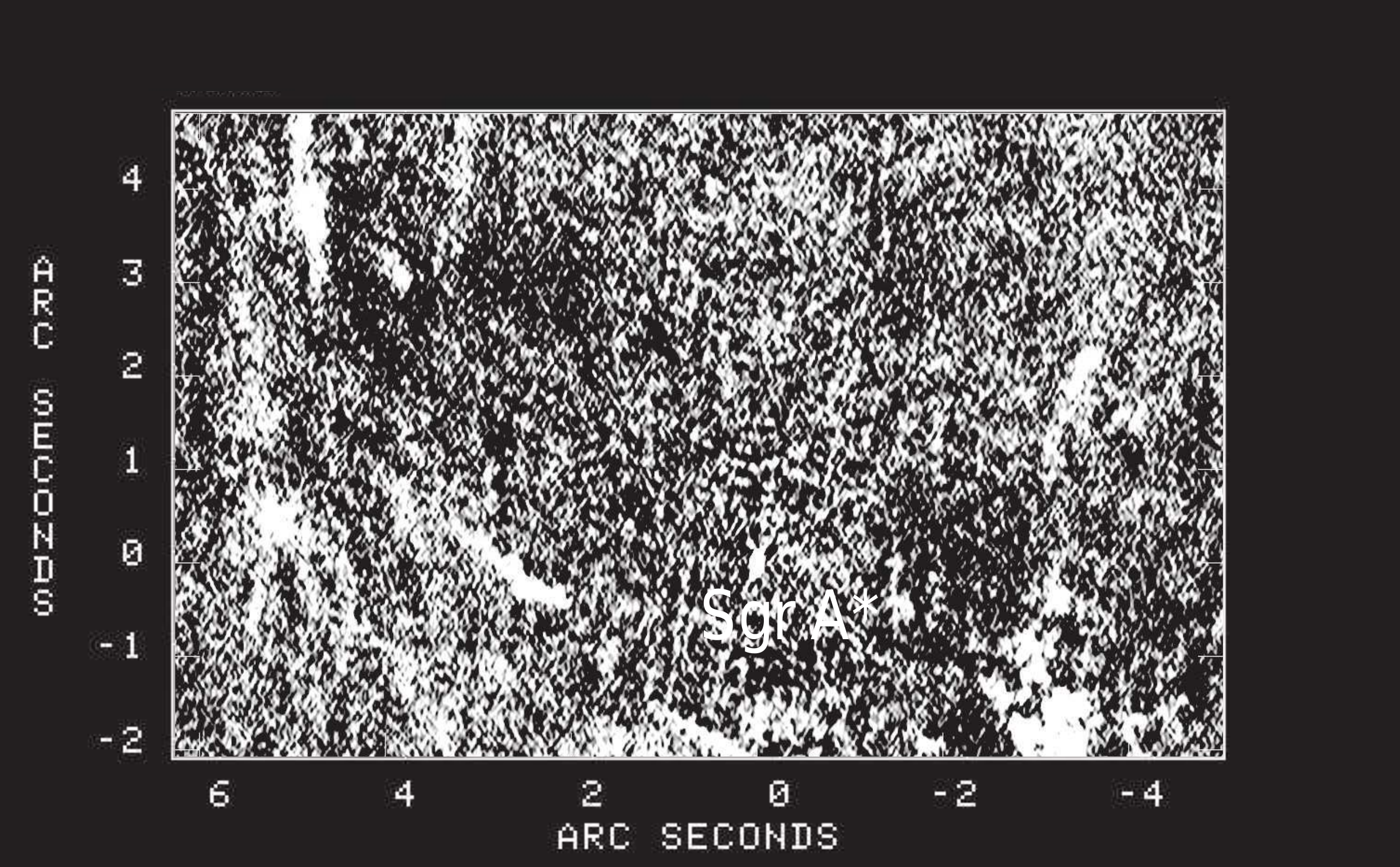} 
\includegraphics[scale=.3,angle=0]{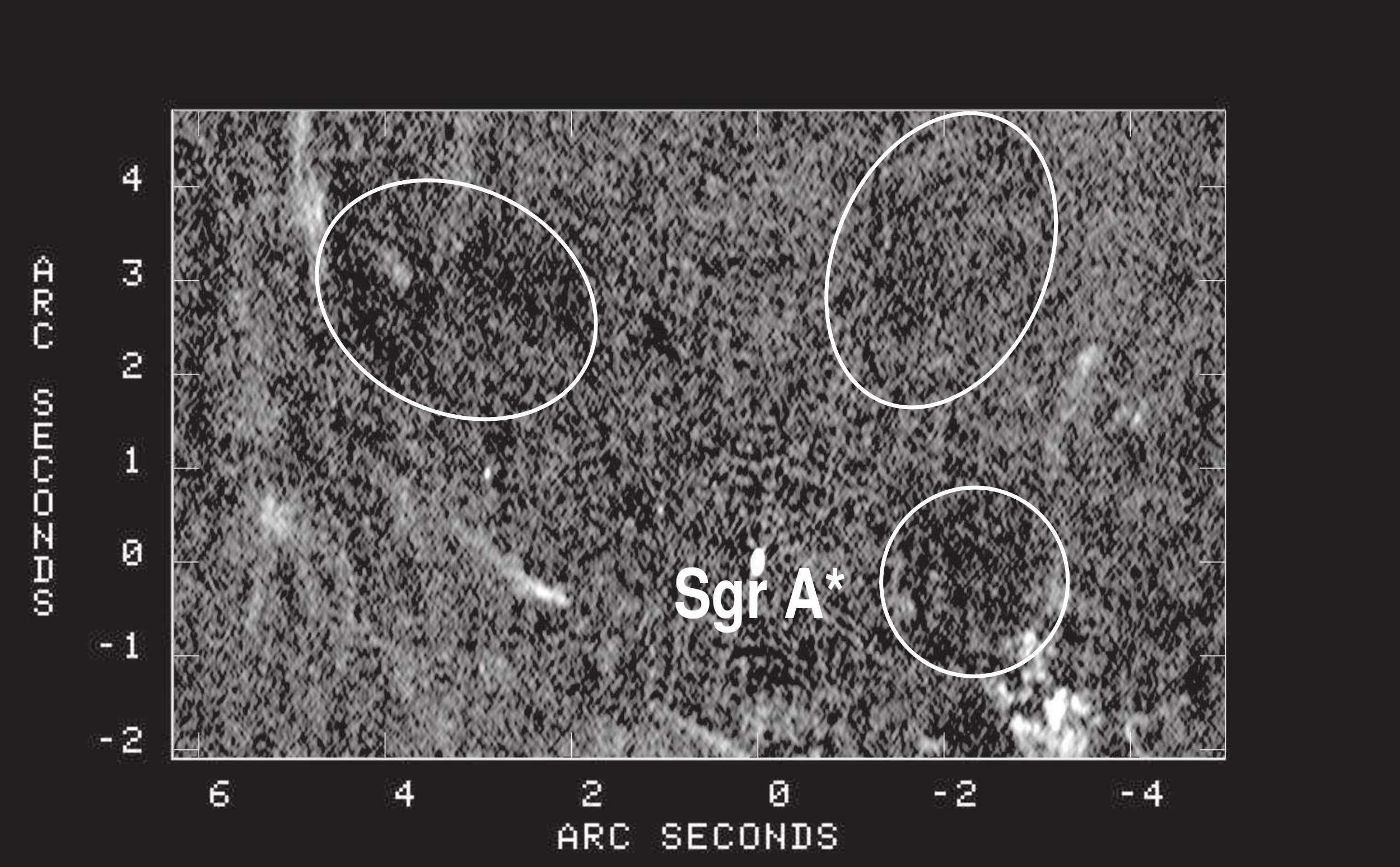} 
\includegraphics[scale=.3,angle=0]{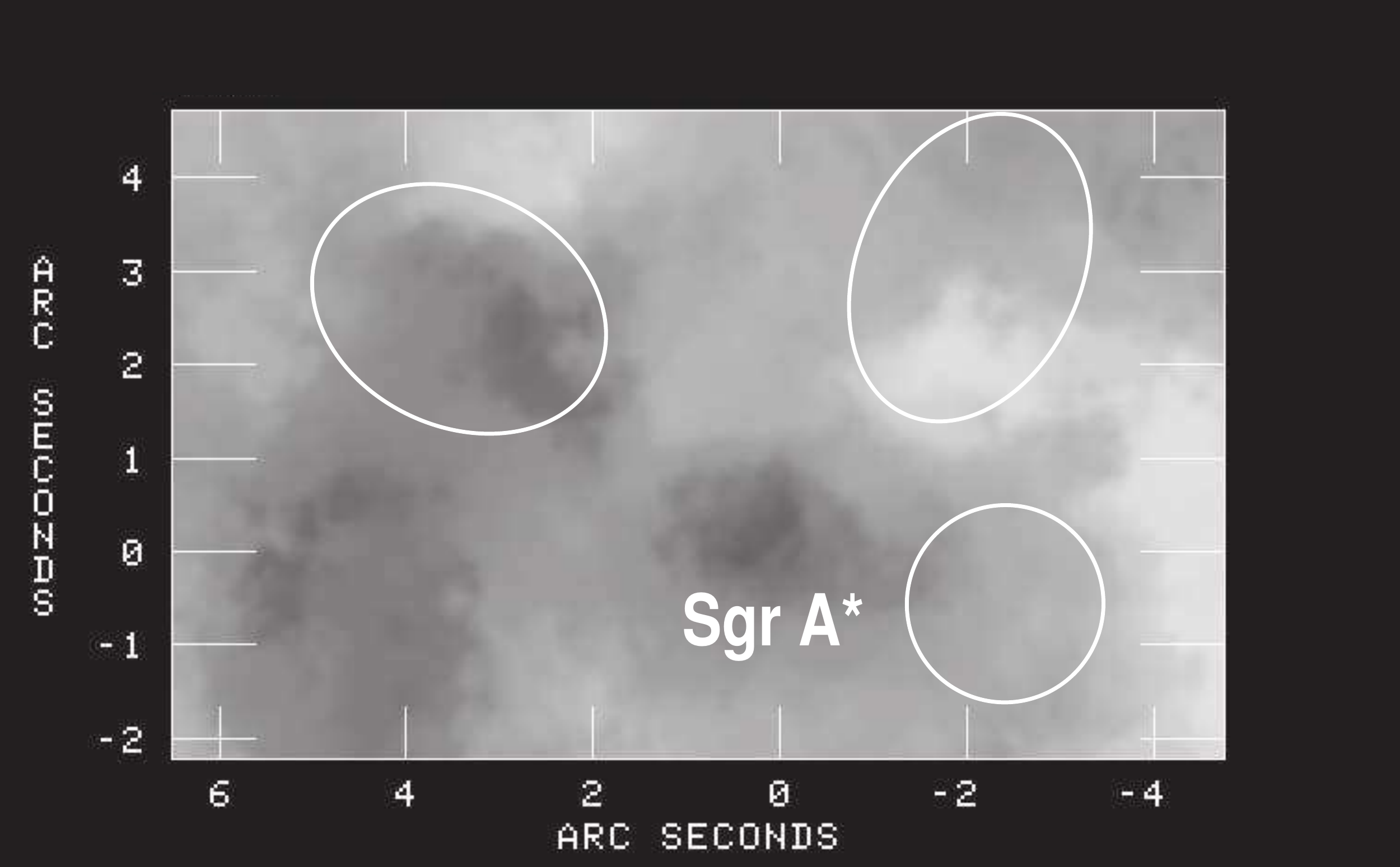} 
\includegraphics[scale=.3,angle=0]{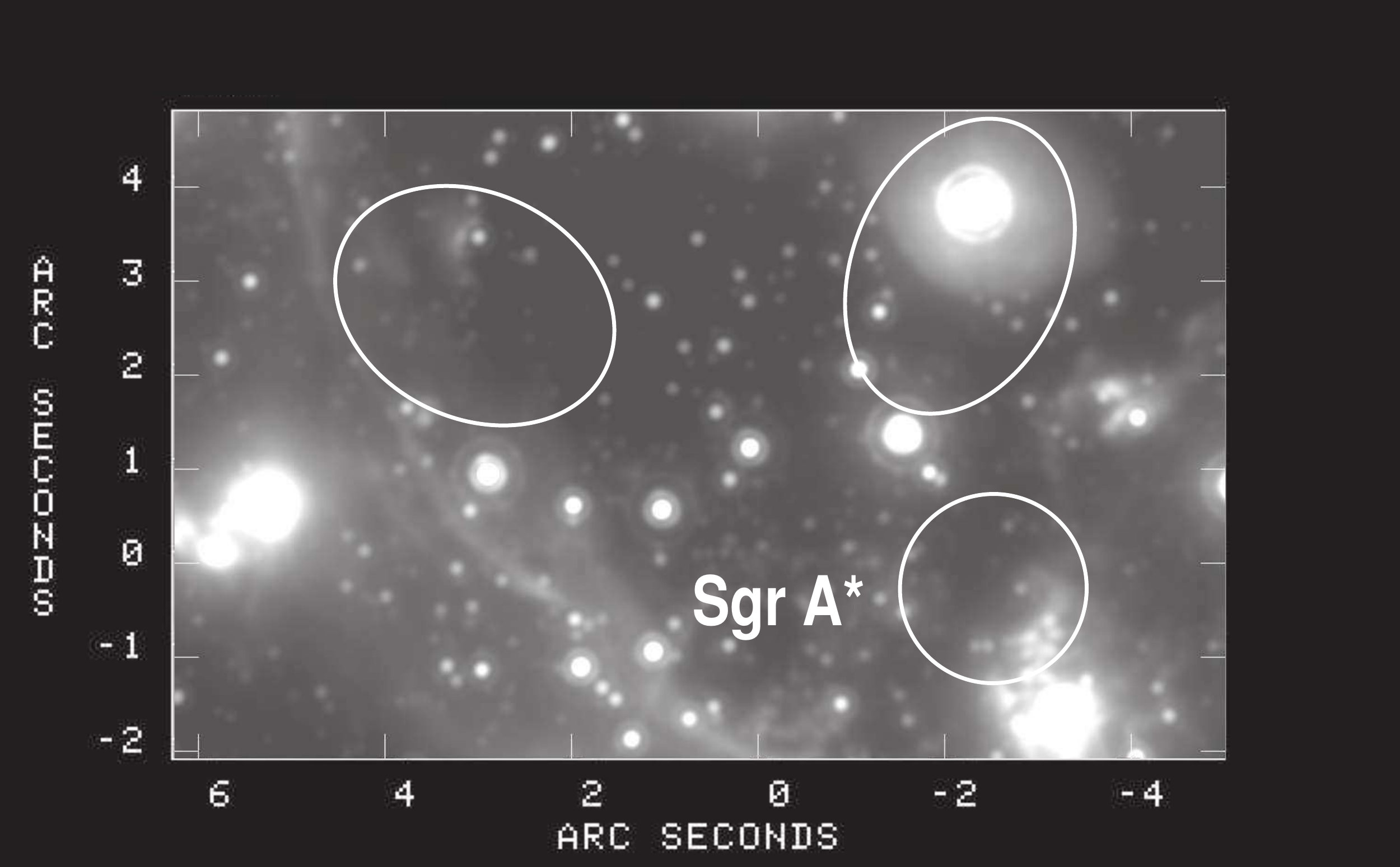} 
\caption{ 
%{\it (a) Top} 
%A grayscale image of the region surrounding IRS 3 and IRS 7 showing dark patches 
%coincident with the envelopes of  these two stars at 44 GHz with 
%at a resolution of $0.084''\times0.042''$ (PA=$-5.65^{\circ}$).  The white dashed lines trace the
%dark radio features. 
%{\it (b) Bottom} Similar to (a) except at 3.8 $\mu$m. 
{\it (a-b)}  The top two  images show the inner 10$''\times7''$ of Sgr A* with different
contrasts in order to bring out the extended radio dark clouds and star associated with IRS 3. 
{\it (c-d)} 
The bottom two figures have the same size as 
(a-b) except they show the extinction  and  3.8 $\mu$m images (Sch\"odel et al. 2009).  
}
\end{figure}
%(grayscale flux -0.1 to 300 mJy beam$^{-1}$). 
%resolution of $10.8''\times5.5''$ (PA=$-2.6^{\circ}$). 

\begin{figure}
\center
\includegraphics[scale=.4,angle=0]{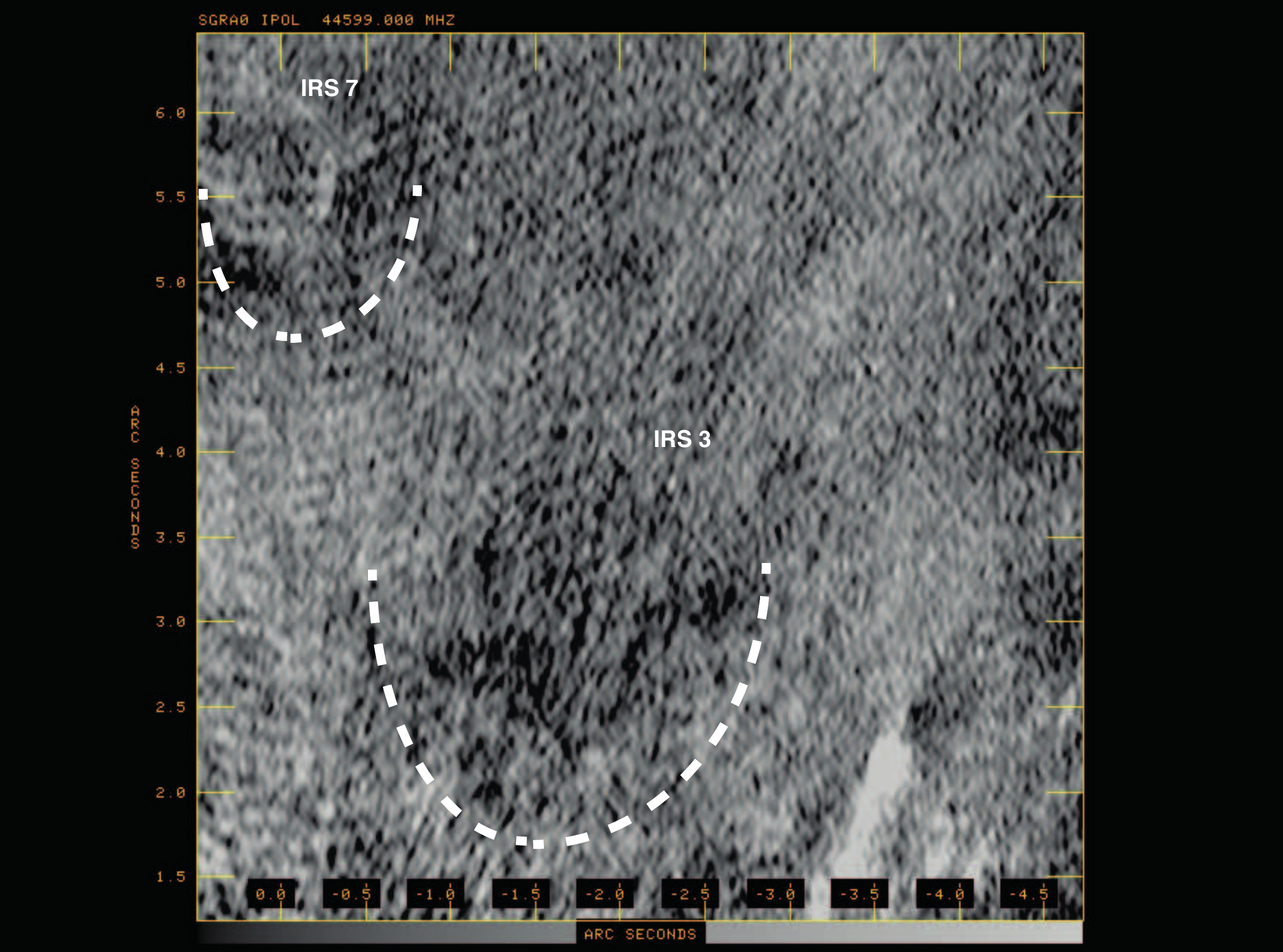}
\includegraphics[scale=.4,angle=0]{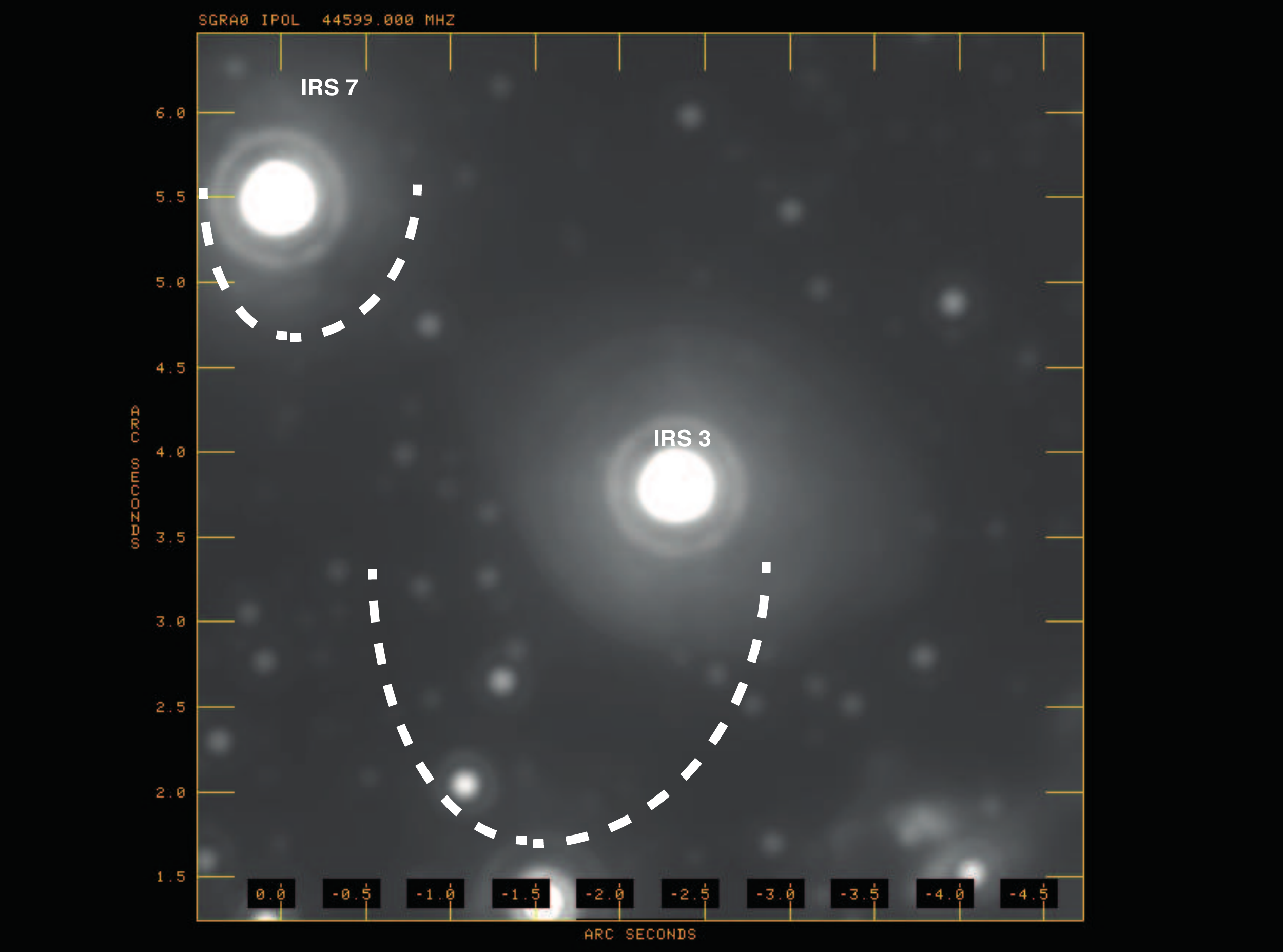}
\caption{
{\it (a) Top} A grayscale image of the region surrounding IRS 3 and IRS 7 showing dark patches
coincident with the envelopes of  these two stars at 44 GHz with
at a resolution of $0.084''\times0.042''$ (PA=$-5.65^{\circ}$).  The white dashed lines trace the
dark radio features.
{\it (b) Bottom} Similar to (a) except at 3.8 $\mu$m.
}
\end{figure}

\begin{figure} 
\includegraphics[scale=0.6,angle=0]{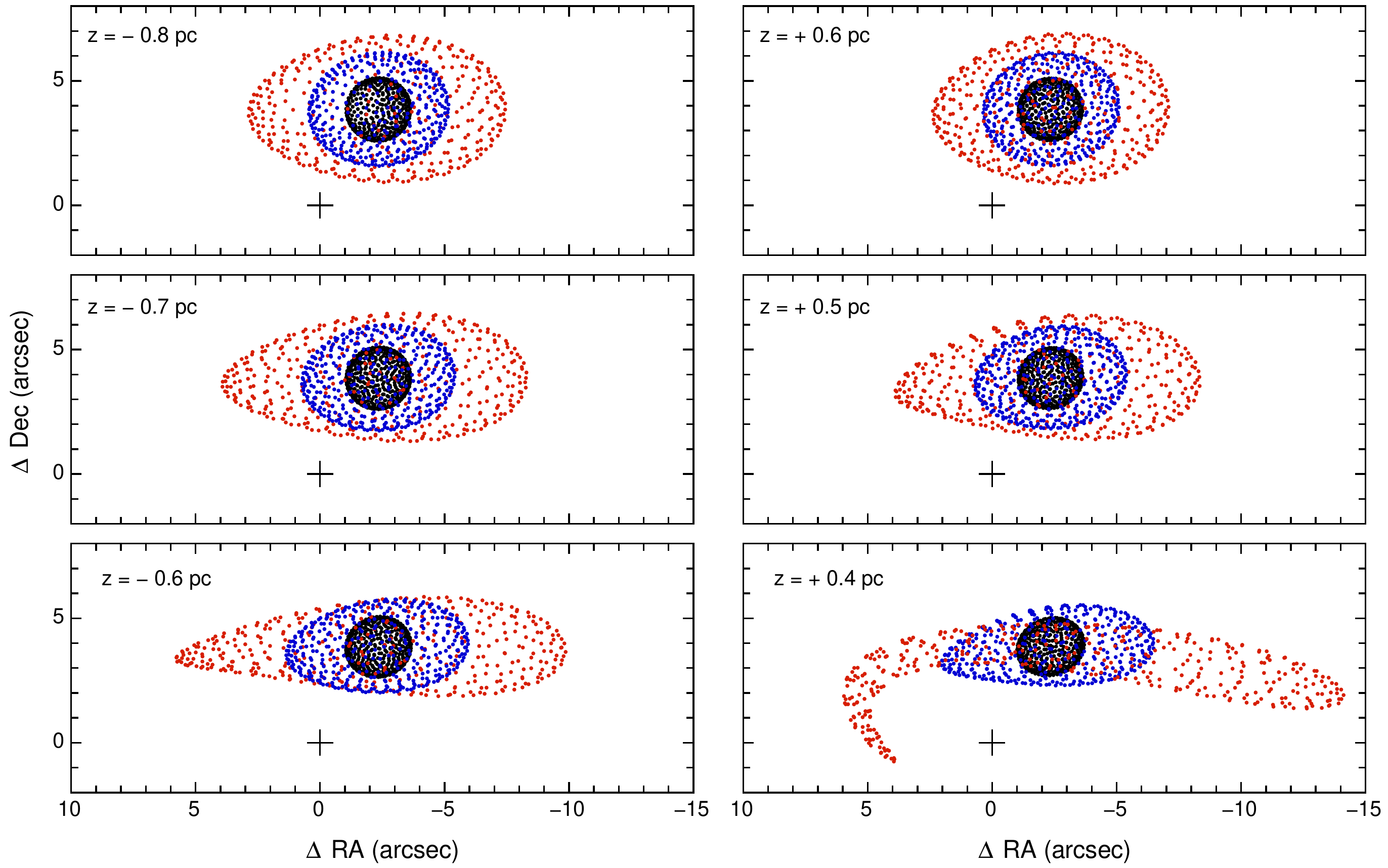} 
\caption{Tidal distortion of the envelope of IRS 3 for different assumed positions along the line of sight, either 
in the foreground (left column), or the background (right column) relative of Sgr A*.  In each panel, black, red, 
and blue points indicate the position in the sky of fluid elements ejected radially from IRS 3 2\,500, 5\,000, and 
7\,500 yrs ago, respectively, at the assumed wind velocity 20\,\kms (see text).  The black cross in each panel 
indicates the location of Sgr A*.}
\end{figure}

\end{document}